\documentstyle[epsfig,10pt] {article}
\begin{document}

\title{
    On the Statistics of Small Scale Turbulence and its Universality}

\author{
    Ch. Renner$^1$, B. Reisner$^1$, St. L\"uck$^1$,  J.~Peinke$^1$ \\
    $^1$ Phys. Institut, EP II, Universit\"at Bayreuth, D-95440 Bayreuth\\
\and R. Friedrich$^2$ \\
    $^2$  Institut f\"ur Theo. Physik und Synergetik,\\
    Universit\"at Stuttgart,  D-70550 Stuttgart}

\maketitle

\begin{abstract}

We present a method how to estimate from experimental data of a 
turbulent velocity field the drift and the diffusion coefficient of a 
Fokker-Planck equation.  It is shown that solutions of this 
Fokker-Planck equation reproduce with high accuracy the statistics of 
velocity increments in the inertial range.  Using solutions with 
different initial conditions at large scales we show that they 
converge.  This can be interpreted as a signature of the universality 
of small scale turbulence in the limit of large inertial ranges.

\end{abstract}

\section{Introduction}

The common picture of fully developed local isotropic turbulence is 
that the velocity field is driven by external fields on large scales.  
By this driving energy is fed into the system at scales larger than 
the integral length $L_0$.  A cascading process will transport this 
energy to smaller and smaller scales until at the viscous length scale 
$\eta$ the injected energy is finally dissipated by viscous effects 
\cite{YMF,Frisch}.  It is commonly believed that this picture of a 
cascade leads to the universality of the statistical laws of small 
scale turbulence.

The standard quantity to study the statistics of turbulent fields is 
the longitudinal velocity increment of the length scale $L_{i}$ 
defined as $ v_i = u(x+L_i) - u(x)$, where $u$ denotes the velocity 
component in the direction of the separation $L_{i}$.  $x$ is a 
selected reference point.

Two common ways have been established to analyze the statistical 
content of the velocity increments.  On one side the structure 
functions have been evaluated by supposing a scaling behavior
\begin{eqnarray}\label{struc}
   <(v_{i})^q> = \int \!\!  dv_{i}\: (v_{i})^q \:P(v_{i},L_i) \propto 
   L_{i}^{\zeta_{q}},
\end{eqnarray}
where $P(v_{i},L_i)$, denotes the probability density function (pdf) 
for $v_i$ at the length scale $L_i$.  Using the recently proposed 
so-called extended selfsimilarity \cite{ESS} it has become possible to 
evaluate the characterizing scaling exponents $\zeta_{q}$ quite 
accurately c.f.  \cite{Arneodo,Sreeni}.  On the other side one is 
interested to parameterize directly the evolution of the probability 
density functions (pdf) $P(v_{i},L_i)$ c.f.  \cite{Benzi,Castaing}.

One major challenge of the research on turbulence is to understand 
small scale intermittency which is manifested in a changing form of 
$P(v_{i},L_i)$ or equivalently in a nonlinear $q$-dependence of the 
scaling exponents $\zeta_{q}$, provided that the scaling assumption is 
valid.

Although it is well known that, due to the statistic of a finite 
number of data points (let say $10^7$ data), it is not possible to 
determine accurately scaling exponents $\zeta_{q}$ for $q>6$ 
\cite{xiq>6}, and that there are different experimental indications 
that no good scaling behavior is present \cite{nonscaling,PRL}, for 
long time the main effort has been put into the understanding of the 
q-dependence of $\zeta_{q}$ (for actual reviews see 
\cite{Frisch,Sreeni}.  That there have been less attempts to analyze 
directly the pdfs may be based on the fact that up to now the scaling 
exponents are regarded as the simplest reduction of the statistical 
content and that this analysis does not depend on model assumptions.  
In contrast to this, proposed parameterizations of the form of the pdfs 
\cite{Benzi,Castaing}, although they are quite accurate, are still 
based on some additional assumptions on the underlying statistics.  
Based on the recent finding that the turbulent cascade obeys a Markov 
process in the variable $L_{i}$ and that intermittency is due to 
multiplicative noise \cite{PRL,PhysD,EuroPL}, we show in Section III 
that it is possible to estimate from the experimental data a 
Fokker-Planck equation, which describes the evolution of the pdfs with 
$L_i$.  We show that this Fokker-Planck equation reproduces accurately 
the experimental probability densities $P(v_{i},L_i)$ within the 
inertial range.  Thus an analysis of experimental data is possible 
which quantifies the statistical process of the turbulent cascade, and 
which neither depends on scaling hypotheses nor on some fitting 
functions \cite{dynamic}.  Having determined the correct Fokker-Planck equation for 
an experimental situation, in section VI we show how solutions of this 
equation with different large scale pdfs will converge to universal 
small scale statistics.  This finding gives evidence for the 
universality of small scale turbulence.

\section{Experimental data}

The results presented here are based on $10^7$ velocity data points.  
Local velocity fluctuations were measured with a hot wire anemometer 
(Dantec Streamline 90N10) and a hot wire probe (55P01) with a spatial 
resolution of about $1 mm$.  The sampling frequency was 8 kHz.  The 
stability of the jet was verified by measurements of the self-similar 
profiles of the mean velocity according to \cite{jet}.  The turbulence 
measurements were performed by placing the probe on the axis of a free 
jet of dry air developing downwards in a closed chamber of the size 2m 
x 1m x1m.  To prevent a disturbing counterflow of the outflowing air, 
an outlet was placed at the bottom of the chamber.  The distance to 
the nozzle was 125 nozzle diameters D. As nozzle we used a convex 
inner profile \cite{BuemFied} with an opening section of $D=8mm$ and 
an area contraction ratio of 40.  Together with a laminarizing 
prechamber we have achieved a highly laminar flow coming out of the 
nozzle.  At a distance of 0.25 D from the nozzle no deviation from a 
rectangular velocity profile was found within the detector resolution.  
Based on a 12bit A/D converter resolution no fluctuations of the 
velocity could be found.  The velocity at the nozzle was $45.5 m/s$ 
corresponding to a Reynolds number of $2.7 \: 10^4$.  At the distance 
of 125 D we measured a mean velocity of $2.25 m/s$, a degree of 
turbulence of 0.17, an integral length $L_{0}=67mm$, a Taylor length 
$\theta= 6,6mm$ (determined according to \cite{Taylorlength}), a 
Kolmogorov length $\eta=0.25mm$, and a Taylor Reynolds number 
$R_{\theta}=190$.

The space dependence of the velocity increments ${v_i}$ was obtained 
by the Taylor hypotheses of frozen turbulence.  For the structure 
functions we found a tendency to scaling behavior for $L_{0} \ge L_{i} 
\ge \theta$.  Intermittency clearly emerge as $L_{i} \rightarrow 
\theta$, as shown in Fig.1 by the different form of the pdfs for 
different scales $L_i$.  
\begin{figure}[ht]
  \begin{center}
     \epsfig{file=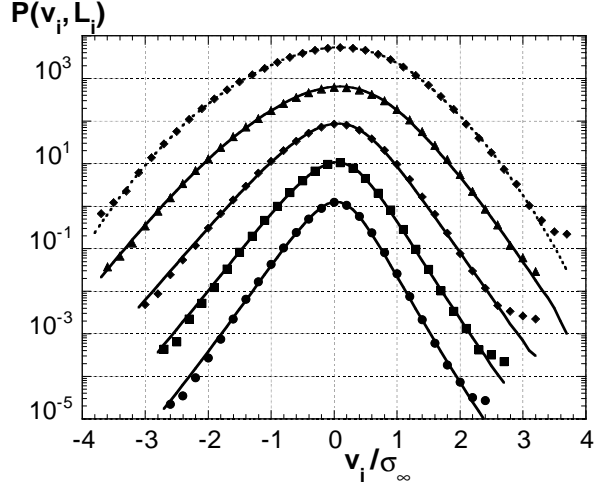, width=8.0cm}
  \end{center}
  \caption{ Probability density functions $P(v,L)$ determined from the 
  experimental data (bold symbols) and calculated pdfs (lines) by the 
  numerical iteration with the experimentally determined Fokker-Planck 
  equation (\ref{FokkerPlanck}) and (\ref{coeff}).  The length scales 
  $\lambda=ln(L_{0}/L)$ were: 0, 0.52, 1.04, 1.56 and 2.24 going from 
  up to down.  For the numerical iteration as initial condition an 
  empirical fitting function for the large scale pdf was used which is 
  shown by a broken line.  For clarity of presentation the pdfs were 
  shifted in y-direction.}

\end{figure}

\section{Measurement of Kramers-Moyal coefficients}

Next, we show how to determine from experimental data appropriate 
statistical equations to characterize the turbulent cascade.  The 
basic quantity for this procedure is to evaluate the cascade by the 
statistical dependence of velocity increments of different length 
scales at the same location $x$ \cite{ZPhysStat}.  Either 
two-increment probabilities $p(v_{2},L_{2};v_{1},L_{1})$ or 
corresponding conditional probabilities $p(v_{2},L_{2}|v_{1},L_{1})$ 
are evaluated from the whole data set.  (Here we use the convention 
that $L_{i+1}<L_i$.)  Investigating the corresponding three-increment 
statistics we could provide evidence that the evolution of these 
statistics with different $L_{i}$ fulfill the Chapman-Kolmogorov 
equation \cite{Risken,PRL,PhysD}.  Furthermore evidences of the 
validity of the Markov process are shown as long as the step size 
between different $L_{i}$ is larger than an Markov length $L_{mar}$, 
which is in the order of some $\eta$ \cite{EuroPL}.  Thus we know 
\cite{Risken} that the evolution of the increments $v_{i}$ is given by 
a master equation without the involvement of memory functions and that 
the evolution of the pdfs $P(v_{i},L_i)$ are described by a 
Kramers-Moyal expansion:
\begin{eqnarray}\label{KrMoyEnt}
   -\frac {d}{d L_{i}} P(v_{i},L_i)= \sum_{k=1}^{\infty} 
   [-\frac{\partial }{\partial v_{i}}]^k D^{(k)}(v_{i},L_{i}) \; 
   P(v_{i},L_i).
\end{eqnarray}
Where the Kramers-Moyal coefficients $D^{(k)}(v_{i},L_{i}) $ are 
defined by the following conditional moments:
\begin{eqnarray}\label{MkDk}
   M^{(k)}(v_{1},L_{1},\delta) & = & \frac{1}{n!} \frac{1}{\delta} 
   \int \!\!  dv_{2} \: (v_{2}-v_{1})^k \: p(v_{2}, L_{2} | v_{1}, 
   L_{1}) 
   \nonumber \\
   D^{(k)}(v_{1},L_{1}) &= & \lim_{\delta \rightarrow 0} 
   M^{(k)}(v_{1},L_{1},\delta) ,
\end{eqnarray}
where $\delta = L_{1}-L_{2}$.  Knowing the Kramers-Moyal coefficients, 
and thus the evolution of $P(v_{i},L_i)$, also the differential 
equations for the structure functions are obtained easily \cite{PRL}
\begin{equation}\label{DGStrFkt}
   - \frac {d} {d L} <(v(L))^q> = 
   \sum_{n=1}^{q-1} \frac {q !} {(q-n)!} <D^{(n)} v^{q-n}> .
\label{MomEq}
\end{equation}
If the Kramers-Moyal coefficients have the form $D^{(q)} = d_{q} 
v^q/L$ (where $d_{q}$ are constants) scaling behavior of (\ref{struc}) 
is guaranteed with $\zeta_{q}= -\sum_{n=1}^{q-1} \frac {q !} {(q-n)!} 
d_{n} $.  For a discussion of this finding in terms of multifractality 
and multiaffinity see \cite{ZNatur}.  The $1/L$ dependence of 
$D^{(q)}$ indicates that the natural space variable is $ln L$.  Thus 
we take in the following as the space variable $\lambda_{i}= 
ln(L_{0}/L_{i})$.  Furthermore we normalize the velocity increments to 
the saturation value of $\sigma_{\infty} = \sqrt{<v^2_{\infty}>}$ for 
length scales larger than the integral length.

An important simplification is achieved if the fourth Kramers-Moyal
coefficient is zero, than the infinite sum of the Kramers-Moyal
expansion (\ref{KrMoyEnt}) reduces to the Fokker-Planck equation of
only two terms \cite{Risken}:
\begin{eqnarray} \label{FokkerPlanck}
   \frac{d} {d \lambda} P(v,\lambda) = 
   \left[ -\frac{\partial}{\partial v} \tilde D^{(1)}(v,\lambda) 
   +\frac{\partial^2}{\partial v^2} \tilde D^{(2)}(v,\lambda) \right] 
   \! P(v,\lambda) .
\end{eqnarray}

Now only the drift term $\tilde D^{(1)}$ and the diffusion term 
$\tilde D^{(2)}$ are determining the statistics completely.  Form the 
corresponding Langevin equation, characterizing the evolution of a 
spatially fixed increment, we know that $\tilde D^{(1)}$ describes the 
deterministic evolution whereas $\tilde D^{(2)}$ describes the noise 
acting on the cascade.  If $\tilde D^{(2)}$ shows a v-dependency one 
speaks of multiplicative noise.

In Fig.  2 we show the evaluated conditional moments $\tilde M^{(1)}$ 
and $\tilde M^{(2)}$ as they evolve for $\delta \lambda \rightarrow 
0$.  For the higher Kramers-Moyal coefficients we found that $\tilde 
M^{(4)} < 10^{-2} \tilde M^{(2)}$.  Thus we take the Fokker-Planck 
equation as the adequate description.
\begin{figure}[ht]
    \begin{center}
	   \epsfig{file=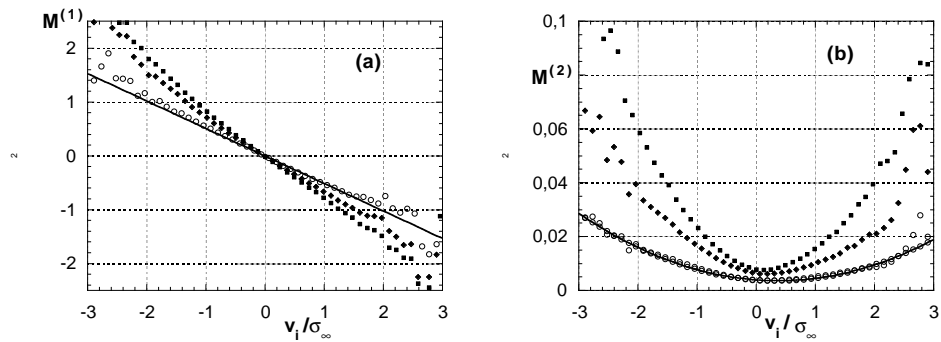, width=12.5cm} 
	\end{center}
	\caption{ The conditional moments $\tilde M^{(1)}(v,\lambda,\delta 
	\lambda)$, (a), and $\tilde M^{(2)}(v,\lambda,\delta \lambda)$, 
	(b), for $\lambda = 1.48$ and $\Delta \lambda $= 0.04 ($\circ$), 
	0.06 and 0.20.  The fits to the experimental data by polynomial 
	functions (see text) are shown by solid lines.  These fits were 
	done only with respect to the interval $-2 < v < 2)$.  }
	
\end{figure}

To estimate the $\tilde D^{(1)}$ and $\tilde D^{(2)}$ coefficient 
properly we have performed the following three procedures based on the 
assumption that we can approximate the Kramers-Moyal coefficient by 
the following polynomials: $\tilde D^{(1)}= \gamma (\lambda)$ and 
$\tilde D^{(2)}= \alpha(\lambda) + \delta(\lambda)*v_i + 
\beta(\lambda)*v_i^2$.  At first, we used the same polynomial forms 
for the $\tilde M$-coefficients and estimated the coefficients of the 
polynomials for $\delta \lambda \rightarrow 0$.  Here problems arise 
due to the finite value of the Markov-length and due to the finite 
resolution of our detector.  Secondly, inserting the polynomials of 
$\tilde D^{(1)}$ and $\tilde D^{(2)}$ into (\ref{DGStrFkt}) we can 
determine directly the coefficients $\alpha$ to $\delta$ from the 
measured structure functions (the highest order of the structure 
function we used was 6).  Problems arise due to the noise of the 
derivative of the structure functions, thus only the magnitude of the 
coefficient could be estimated.  Thirdly, we have evaluated 
analogously the corresponding structure functions of 
$P_{even}(v,L)=1/2(P(v,L)+P(-v,L))$ and 
$P_{odd}(v,L)=1/2(P(v,L)-P(-v,L))$.  \cite{comment}

From all these estimations we got a good guess of the values of 
$\alpha(\lambda)$ to $\delta(\lambda)$.  The best results we obtained 
for $\gamma$.  The worst result was obtained for $\beta$, for which we 
finally used the value $\mu/18$ expected from the Kolmogorov picture of 
intermittency \cite{KO62,PRL}.  As the final values we got:
\begin{eqnarray}\label{coeff}
   \gamma(\lambda) & = & 
   0.36 exp \left \{ (\frac{\lambda -0.68} {2.1} )^3 \right \} \nonumber \\
   \alpha(\lambda) & = & 0.02 exp \left \{ - \frac{\lambda} {1.3}  \right \}
   \nonumber \\
   \delta(\lambda) & = & 0.013 exp ( -1.2 \lambda ) \nonumber \\
   \beta & = & 0.017 \qquad .
\end{eqnarray}
Next, let us briefly comment this result.  For $\gamma$ we find that 
it has a constant value of 1/3 in the inertial range ($0<\lambda<2$).  
The functional dependence of $\gamma$ is given to extend the $\lambda$ 
dependence into the dissipation range.  The values $\delta$ and 
$\alpha$ are violating scaling behavior.  Both, $\alpha$ and 
$\delta$, are decaying exponentially with $\lambda$.  For scales 
larger than the integral length these two coefficients are large and 
are responsible for the building up of the skewness, because these 
terms allow the change of the sign of the velocity increments within 
the stochastic process.

As further test of the validity of our approach we have calculated the 
evolution of the pdfs by a numerical iteration with the Fokker-Planck 
equation.  As an initial condition the pdf $P(v_0,L_{0})$ at the 
integral scale was approximated by an empirical function and then 
inserted into the numerical iteration.  We found that the evolution of 
the pdfs depends sensitively on the chosen coefficients.  (Finally, we 
changed the numbers in (\ref{coeaff}) in the range of some percent to 
obtain a best result.)  The result of the numerical iteration of the 
Fokker-Planck equation with the above mentioned coefficients are shown 
in Fig.  1 by solid lines.

\section{Universality}

Let us now discuss how the turbulent cascade will be affected, if the 
statistics on large scales is changed as it may be the case for 
different driving forces.  As an extreme case we have chosen a box like 
form for a large scale pdf (the standard deviation and the skewness 
was adapted to the one of our experimental pdf for $L_{0}$ (see Fig.  
3)).  Next, using the above mentioned coefficients (\ref{coeff}), we 
have iterated simultaneously the experimental approximation of 
$P(v_0,L_0)$, see Fig.  1, and the box like pdf.  As shown in Fig.3 we 
see that under the iteration the synthetic pdf becomes more and more 
similar to the experimental ones.  This convergence has been 
quantified by the $\chi^2$ measure ($\chi^2 \propto \sum (P_1 - 
P_2)^2/P_1$ ).  We find that $\chi^2$ decays about exponentially with 
$\lambda$.  Thus, we see that for a sufficient large cascade any large 
scale pdf will converge against universal small scale pdfs.
\begin{figure}[ht]
  \begin{center}
    \epsfig{file=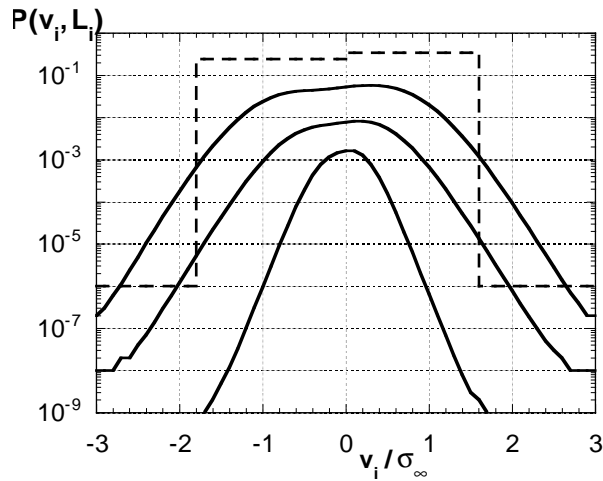, width=8.0cm}
  \end{center}
\caption{Evolution of a synthetic box-shaped pdf (broken line) under 
the numerical integration with the Fokker-Planck equation using the 
coefficients of (\ref{coeff}).  From top to bottom $\lambda =$ 0, 1.4, 
2.2 and 3.0.  Presentation as in Fig.1.  }
\end{figure}
\section{Conclusion}

We have presented further evidence that the statistics of fully 
developed turbulence is based on a Markov process in the space variable 
$L$.  It has been shown how the drift coefficient and the diffusion 
coefficient of a describing Fokker-Planck equation can be determined 
from experimental data.  With these coefficients it is possible to 
calculate accurately the pdfs of a cascade.  As a further result we 
have shown that situations where the large scale statistics differ, 
for example due to different driving forces, will show the same small 
scale statistics, as long as the same Fokker-Planck equations apply.  
Therefore, universality of small scale turbulence can best be 
characterized by comparing the drift and diffusion coefficients.

The next challenging questions are, how do these result depend on the 
Reynolds number and what happen in anisotropic turbulence.  For the 
ladder point recently evidence has been found that for sufficiently 
small scales again the results of local isotropic turbulence hold 
\cite{inhomogeneous}.  We believe that these findings do not only play 
an important role for the complete statistical characterization of 
turbulence but also should have quite practical importance for 
numerical simulations of flow situations with high Reynolds numbers.

\section{Acknowledgments}

We want to acknowledge discussions with A. Naert, P. Talkner, K. 
Zeile, H. Brand.  J.P. acknowledges the financial support of the 
Deutsche Forschungsgemeinschaft.


\begin{thebibliography}{99}
\bibitem{comment} The Fokker-Planck equations for $P_{even}$ and 
$P_{odd}$ have the same form as (\ref{FokkerPlanck}), only if the 
coefficient $\delta =0$, this may clear up the question of the sense 
of evaluating moments of the absolute values of $v_i$, see for example 
footnote on page 446 in \cite{Sreeni}.
\end{thebibliography}
\end{document}